# EFFICIENT OPTIMISATION OF STRUCTURES USING TABU SEARCH

Andy M. Connor†, Keith A. Seffen‡, Geoffrey T. Parks† and P. John Clarkson†

†Engineering Design Centre, Department of Engineering
University of Cambridge, Trumpington Street, Cambridge, CB2 1PZ, UK

‡Department of Mechanical Engineering
UMIST, PO Box 88, Manchester, M60 1QD, UK

E-mail: amc50@eng.cam.ac.uk, keith.seffen@umist.ac.uk, gtp@eng.cam.ac.uk, pjc10@eng.cam.ac.uk

## ABSTRACT

This paper presents a novel approach to the optimisation of structures using a Tabu search (TS) method. TS is a metaheuristic which is used to guide local search methods towards a globally optimal solution by using flexible memory cycles of differing time spans. Results are presented for the well established ten bar truss problem and compared to results published in the literature. In the first example a truss is optimised to minimise mass and the results compared to results obtained using an alternative TS implementation. In the second example, the problem has multiple objectives that are compounded into a single objective function value using game theory. In general the results demonstrate that the TS method is capable of solving structural optimisation problems at least as efficiently as other numerical optimisation approaches.

## INTRODUCTION

Global optimisation algorithms such as Genetic Algorithms (GAs) and Simulated Annealing (SA) have attracted considerable attention in recent years from not only the structural engineering community but also researchers in many other diverse fields. Although GAs and SA have been shown to be capable of solving optimisation problems that are otherwise intractable, there is a growing interest in other "heuristic" algorithms such as TS. This is fuelled by claims that TS is considerably more effective than other methods (Sinclair, 1994; Borup and Parkinson, 1993). This paper presents results obtained by applying a variable step size TS algorithm that was originally developed for application to parameter sizing in hydraulic circuits (Connor and Tilley, 1997, 1998a, 1998b) which is continuing to evolve in structure and to be applied to other problem domains (Leonard and Connor, 1999).

The TS algorithm presented here differs from many previously published algorithms (Bland, 1994; Fanni et al., 1998) in that the control parameters associated with the search are generally assigned to values that promote computational efficiency rather than algorithmic effectiveness and yet the results achieved in most domains are of high quality. This can be attributed to the unique control algorithm and the use of a variable step size.

## TABU SEARCH

TS (Glover and Laguna, 1997) is a metaheuristic which is used to guide optimisation algorithms in the search for a globally optimal solution. The TS algorithm uses flexible memory cycles of differing time spans to force the search out of local optima and to provide strategic control of how the search progresses through the solution space.

### Short Term Memory

The most simple implementation of TS is based around the use of a hill climbing algorithm. Once the method has located a locally optimal solution the use of the short term memory, or tabu restrictions, ensures that the search does not return to the optimum after the algorithm forces the search out in a new direction. In the TS implementation used in this work, the short term memory contains a list of the last $n$ visited solutions and these are classed as tabu.

The effect of this concept can be illustrated by considering the diagram shown in Figure 1. This shows a contour plot of a two-dimensional function which contains one local and one global optimum and the aim of the search is to find the location with the lowest value.

From the indicated start position the local search algorithm quickly locates the locally optimal solution without the tabu restrictions being considered as a continuous descent is possible. However, when the search reaches the local optimum the aggressive nature of the TS forces the algorithm out of the optimum in the direction that increases the objective function by the smallest amount. Because the last $n$ visited solutions are classed as tabu, the search cannot leave the optimum along the reverse trajectory from which it entered and once it has left the optimum it cannot enter it again. The algorithm therefore forces the search to climb out of the local optimum and in due course it successfully locates the global optimum.

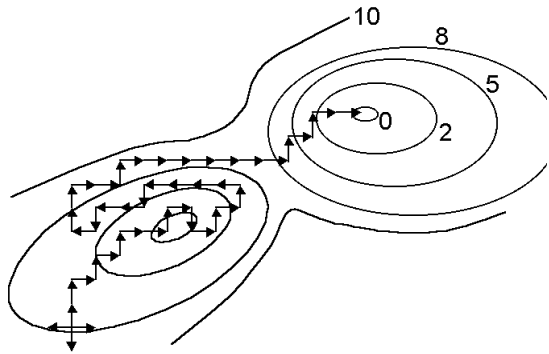

Figure 1. Action of tabu restrictions

Search Intensification and Diversification

The TS short term memory enables the method to leave locally optimal solutions in the quest for the global optimum of a function. However, short term memory alone does not ensure that the search will be both efficient and effective. Search *intensification* and *diversification* techniques are often used first to focus the search in particular areas and then to expand the search to new areas of the solution space. This is normally achieved by the use of longer term memory cycles.

Intermediate and long term memory cycles generally use similar lists of previously visited solutions to guide the search. In the specific implementation used in this work the intermediate term memory cycle is based on a list of the $m$ best solutions found so far. This list is therefore only updated when a new improved solution is found as opposed to whenever a move is made. At certain stages throughout the search process a degree of intensification is achieved by reinitialising the search at a new point generated by considering similarities between the solutions contained in the intermediate memory list.

In the implementation used in this paper, diversification is achieved by using a simple random refreshment although more strategic diversification could be implemented through the use of long term memory.

Hill Climbing Algorithm

The underlying hill climbing algorithm used in this work is based upon the method developed by Hooke and Jeeves (1961). This method consists of two stages, the first of which carries out an initial exploration around a given base point. When a move to a new point (the exploration point) which improves the objective function is identified, the search is extended along the same vector by a factor $k$. This is known as a pattern move. If this new solution has a better objective value than the exploration point, then this point is used as the new base point and the search is repeated. Otherwise, the search is repeated using the exploration point as the new base point.

The algorithm used in this implementation of TS differs in several ways from the standard Hooke and Jeeves algorithm. In the Hooke and Jeeves algorithm each parameter is varied in turn and the *first* move that results in a better objective function value is selected. The implication of this is that not all potential moves are evaluated. In the TS implementation all trial moves are investigated and the *best* move that is not tabu is chosen.

The second difference concerns how the step size is periodically reduced. In the Hooke and Jeeves search when a point is reached from which no improvement can be found then the step size is reduced by a factor of two. In the TS implementation this is not practical as the TS metaheuristic forces the search point out of local optima. The step size is therefore reduced when other conditions apply. A counter is maintained of the number of search moves that have elapsed since an improved solution was found. When this reaches a given value then the search carries out an intensification action. If an improved solution is found then the counter is reset. If no improvement is found then the search continues until the counter reaches a higher preset value at which point diversification is carried out. Again, the search continues and if no improvement is found before the number of moves reaches the next preset level then the step size is reduced. It is worth noting that the number of iterations before intensification occurs is very low (4 iterations) when compared to other implementations that allow up to 150 moves before intensification occurs (Fanni *et al.*, 1998). The entire control algorithm for our TS implementation is shown in Figure 2.

The actual reduction in step size combines aspects of the geometric reduction as used in the original algorithm and aspects of discrete search used in previous work (Connor and Tilley, 1998b). Once the step size is reduced by a factor of two the resulting value is truncated to a multiple of the minimum step size. This ensures that sensible parameter values are used. This approach has been used in other applications (Leonard and Connor, 1999) and is much more efficient than the original linear reduction scheme (Connor and Tilley, 1998b).

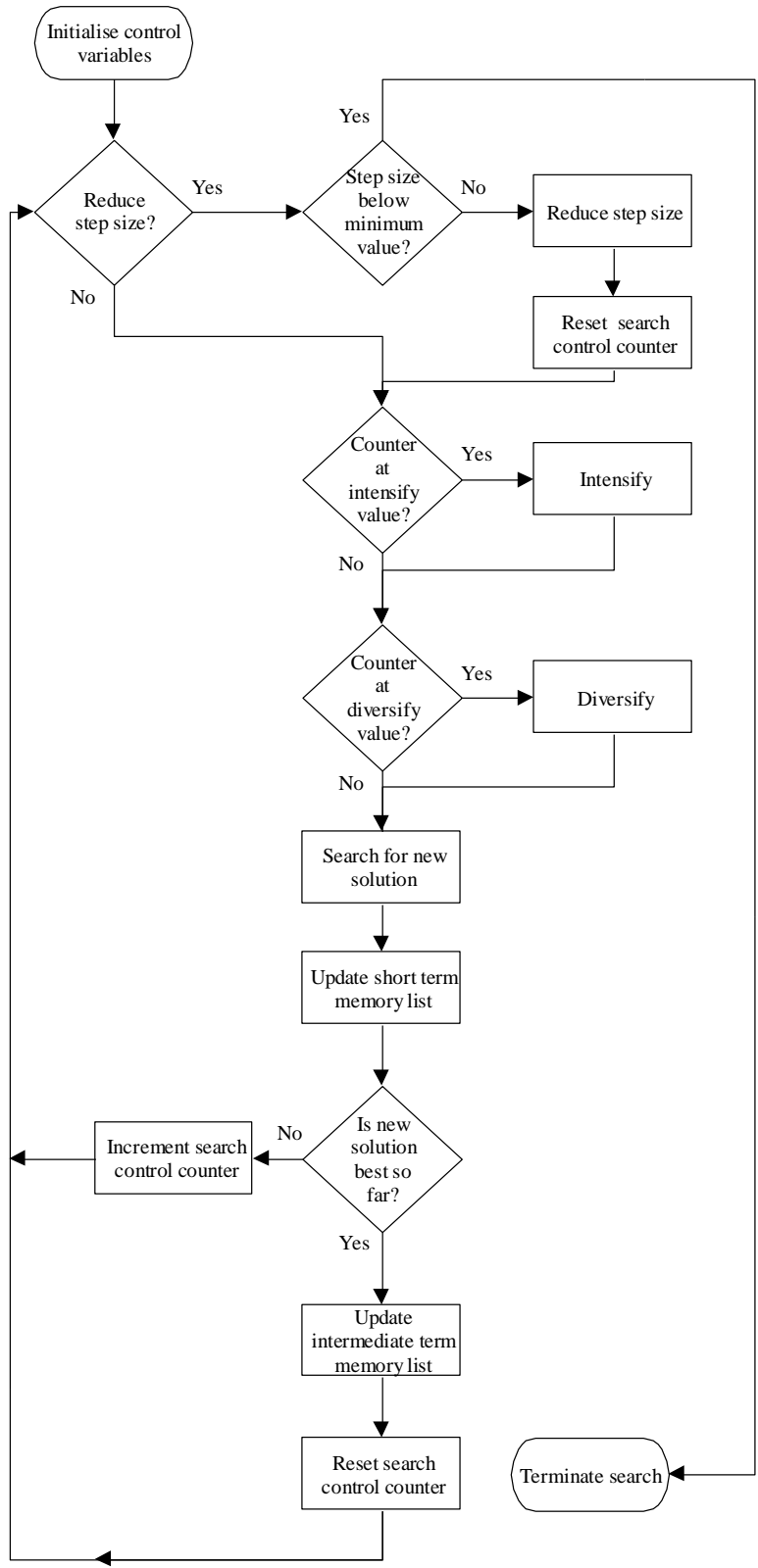

Figure 2. TS control algorithm

## TEN BAR TRUSS EXAMPLE PROBLEM

The structural optimisation problem considered in this paper is the ten bar truss. This example is quite common in the literature (Rajeev and Krishnammorthy, 1992; Bennage and Dhingra, 1995; Bland, 1994) which allows comparisons to be made to previously published work to assess the performance of the TS algorithm. This section will describe the general nature of the problem and later sections will discuss the particular instances from the literature to which the TS performance will be compared. The truss is shown in Figure 3.

The truss is idealised as a set of pin pointed bars connected together at the indicated nodes. The design variables for this problem are the cross-sectional areas of the ten bars in the cantilever truss. The length of each bar is fixed and the externally loaded nodes are as shown in Figure 3.

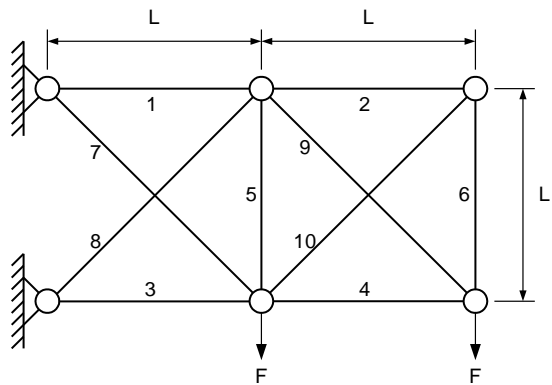

Figure 3. The ten bar truss

The potential design objectives are the minimisation of mass, maximisation of natural frequency and minimisation of total displacement. There are constraints on the maximum allowable stress and the displacement of individual nodes. Due to its statically indeterminate nature, a finite element approach has been used to analyse the structure. However, because of the simplicity of the structure the global mass and stiffness matrices for a linear response were set up by hand using a procedure similar to that of Paz (1985). This enables the mass, deflections and fundamental frequency to be calculated without the use of a commercial finite element package which would have effected the overall computational efficiency.

Application of Tabu Search for Minimum Weight Structure

The current TS algorithm can be applied to the ten bar truss problem with the aim of minimising only the mass of the structure. Both the approach and the results obtained can be compared to a previously published application of TS to this problem (Bland, 1994). In this previously published work, the optimisation was carried out under a number of different loading and constraint conditions. These included additional loading due to the self-weight of the structure and also the inclusion of a buckling constraint for bars in compression. While these refinements result in a more realistic structural model, they do not change the fundamental nature of the optimisation problem, and therefore we restrict ourselves to the study of just one optimisation case.

There are a number of clear weaknesses to the TS algorithm used by Bland. First, the search is run for a fixed number of trials. This seriously limits the implementation as TS is intended to track between large numbers of local optima in order to find the global optimum. However, this limitation is not apparent in Bland's results as the search is initialised from a solution that is very near to the final optimum solution.

Second, the size of the short term memory, or tabu list, is equal in size to the number of potential trials. The effect of this is that no trial solution can be revisited. In itself, this is not a bad strategy but it has serious implications on computational efficiency and system requirements. It is also inefficient as the TS search vector may have to follow very long paths to go from one part of the search space to another if it can *never* revisit solutions.

The constraints and limits for the design variables for the problem solved here are shown in Table 1. This is the simplest case presented by Bland (1994) in which buckling constraints and loading due to self-weight are not included.

| Maximum stress | $\sigma_{max} = 0.16 \times 10^6$ kPa |
| --- | --- |
| Maximum nodal displacement | $\delta_{max} = 0.015$ m |
| Minimum cross-sectional area | $A_{min} = 0.168$ m$^2$ |
| Maximum cross-sectional area | $A_{max} = 0.495$ m$^2$ |

Table 1. Problem constraints

The truss is to be optimised using steel as the material. The properties of the material and structure are given in Table 2.

| Young's modulus | $E = 2.07 \times 10^8$ kN/m$^2$ |
| --- | --- |
| Density | $\rho = 7850$ kg/m$^3$ |
| Length | $L = 3$ m |
| Load | $F = 500$ kN |

Table 2. Material properties and problem definition

Figure 4 shows a plot of convergence of the objective function using the current TS implementation on this problem when it is started from an initial solution for which each cross-sectional area $A_i = 0.761$m$^2$. This is the smallest value to which all the design parameters may be set while allowing the solution to remain feasible (Bland, 1994).

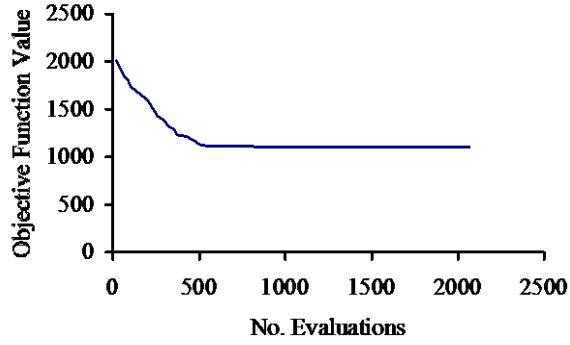

Figure 4. Convergence of the objective function

Table 3 compares the final solution obtained with the solution found by Bland (1994).

|  | TS | Bland |
|---|---|---|
| $A_1$ | 1.022 m² | 0.761 m² |
| $A_2$ | 0.168 m² | 0.268 m² |
| $A_3$ | 0.601 m² | 0.761 m² |
| $A_4$ | 0.341 m² | 0.363 m² |
| $A_5$ | 0.168 m² | 0.168 m² |
| $A_6$ | 0.168 m² | 0.168 m² |
| $A_7$ | 0.361 m² | 0.418 m² |
| $A_8$ | 0.679 m² | 0.646 m² |
| $A_9$ | 0.361 m² | 0.418 m² |
| $A_{10}$ | 0.168 m² | 0.168 m² |
| Mass | 1103.8 kg | 1112.1 kg |

Table 3. Comparison of solutions

It can be seen that the current TS implementation locates a solution with a better objective function value than the solution found by Bland (1994) when started from the same point. The two solutions have some similarities but are obviously distinct. Our TS method is quickly locating solutions with low objective function values (objective function value ≈ 1100 kg after 500 evaluations). Two design variables in Bland's best solution have the same values as those with which they were initialised, perhaps suggesting that the method is not fully exploring the solution space.

Multiobjective Optimisation using a Game Theory Approach

In this example the aim is to minimise not only mass, but also the total nodal deflections as well as maximising the truss natural frequency. These three objectives are compounded into a single objective function by using a game theory approach (Bennage and Dhingra, 1993).

| Maximum stress | $\sigma_{max}$ = 0.25×10⁵ psi |
| --- | --- |
|  | $\sigma_{max}$ ≈ 1.724×10⁵ kPa |
| Maximum nodal displacement | $\delta_{max}$ = 2 in |
|  | $\delta_{max}$ ≈ 0.054 m |
| Minimum cross-sectional area | $A_{min}$ = 0.1 in² |
|  | $A_{min}$ ≈ 6.45×10⁻⁵ m² |
| Maximum cross-sectional area | $A_{max}$ = 33.5 in² |
|  | $A_{max}$ ≈ 2.16×10⁻² m² |

Table 4. Problem constraints

The constraints and limits for the design variables are shown in Table 4 in both imperial and metric units. Imperial units have been included so that a direct comparison to the work of Bennage and Dhingra (1993) can be made.

The three objective functions are calculated using the following equations. The objective for maximising the natural frequency has been transformed into a minimisation problem using a sign convention:

$$f_1 = \sum_{i=1}^{10} \rho A_i l_i \quad (1)$$

$$f_2 = -\omega_1 \quad (2)$$

$$f_3 = \sum_{i=1}^{4} \{\delta_{ix}^2 + \delta_{iy}^2\}^{1/2} \quad (3)$$

Given the values for each objective when the problem is considered as a single objective problem (Bennage and Dhingra, 1993) it is possible to formulate a compounded single objective in the following manner.

$$f = \prod_{i=1}^{3} \{f_{iw} - f_i(X)\} \quad (4)$$

In this equation, $f_{iw}$ is the worst value for the $i$-th objective from the three sets of values obtained from a single objective optimisation and $f_i(X)$ is the value for the $i$-th objective for the current set of design parameters.

By normalising the values between the range of best and worst objective function values from the uni-objective solutions this produces three values between 0 and 1, the product of which measures the performance relative to the least desirable solution. Maximising this function therefore leads to an optimal trade off. The normalisation is carried out using the objective function values of the single objective solutions obtained by Bennage and Dhingra (1993). The truss is to be optimised using aluminium as the material. The properties of the material and structure are given in Table 5.

| Young's modulus | $E = 10^7$ psi |
| --- | --- |
|  | $E \approx 6.895 \times 10^7$ kPa |
| Density | $\rho = 0.1$ lb/in$^3$ |
|  | $\rho \approx 2746.1$ kg/m$^3$ |
| Length | $L = 360$ in |
|  | $L \approx 9.144$ m |
| Load | $F = 100$ klbf |
|  | $F \approx 444.84$ kN |

Table 5. Material properties and problem definition

Figure 5 shows a plot of convergence of the objective function using the current TS implementation on this problem when it is started from the same initial solution as that used by Bennage and Dhingra (1993).

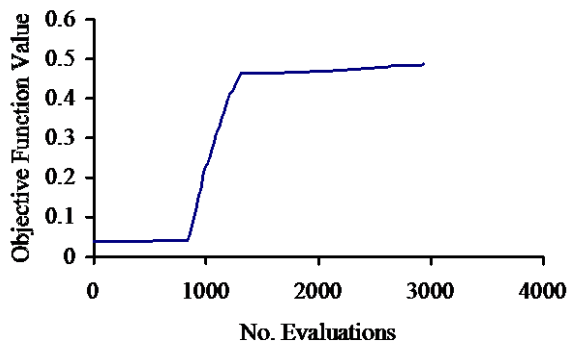

Figure 5. Convergence of the objective function

Table 6 compares the final solution obtained with the solution found by Bennage and Dhingra (1993) and Table 7 compares the values for the different objectives.

The objective function values shown for the Bennage and Dhingra (1993) solution are taken directly from their paper. Using the current model of the truss, slight differences in numerical precision are observed and the actual optimal solution proposed by Bennage and Dhingra violates a constraint on nodal deflections. As the single objective solutions have been used without verifying adjustment during the normalisation it is difficult to make a precise direct comparison. However, it is clear that, after a relatively low number of evaluations, our TS method is locating a high quality solution that has desirable values for the individual objectives. Compared to the solution found by Bennage and Dhingra the solution found by the TS algorithm has improved values for two of the three objectives. However, for the natural frequency objective the value is slightly worse. If these values are considered from a true multiobjective perspective, the solution found by the TS algorithm is pareto optimal. Bennage and Dhingra do not explicitly state how many evaluations their Simulated Annealing based method required to find the solution they give, but it is implied that several thousand were needed. Thus the efficiency of our TS method is certainly comparable and probably superior.

|  | TS | B&D |
| --- | --- | --- |
| $A_1$ | 33.5 in$^2$ | 33.4896 in$^2$ |
|  | ~2.16×10$^{-2}$ m$^2$ | ~2.16×10$^{-2}$ m$^2$ |
| $A_2$ | 1.25 in$^2$ | 1.4392 in$^2$ |
|  | ~8.06×10$^{-4}$ m$^2$ | ~9.28×10$^{-4}$ m$^2$ |
| $A_3$ | 33.5 in$^2$ | 33.4996 in$^2$ |
|  | ~2.16×10$^{-2}$ m$^2$ | ~2.16×10$^{-2}$ m$^2$ |
| $A_4$ | 10.55 in$^2$ | 11.1137 in$^2$ |
|  | ~6.81×10$^{-3}$ m$^2$ | ~7.17×10$^{-3}$ m$^2$ |
| $A_5$ | 1.8 in$^2$ | 1.3353 in$^2$ |
|  | ~1.16×10$^{-3}$ m$^2$ | ~8.61×10$^{-4}$ m$^2$ |
| $A_6$ | 0.1 in$^2$ | 0.1002 in$^2$ |
|  | ~6.54×10$^{-5}$ m$^2$ | ~6.58×10$^{-5}$ m$^2$ |
| $A_7$ | 32.3 in$^2$ | 32.8076 in$^2$ |
|  | ~2.08×10$^{-2}$ m$^2$ | ~2.11×10$^{-2}$ m$^2$ |
| $A_8$ | 32.5 in$^2$ | 33.4843 in$^2$ |
|  | ~2.09×10$^{-2}$ m$^2$ | ~2.16×10$^{-2}$ m$^2$ |
| $A_9$ | 14.0 in$^2$ | 13.2201 in$^2$ |
|  | ~9.03×10$^{-3}$ m$^2$ | ~8.53×10$^{-3}$ m$^2$ |
| $A_{10}$ | 1.85 in$^2$ | 1.9814 in$^2$ |
|  | ~1.19×10$^{-3}$ m$^2$ | ~1.28×10$^{-3}$ m$^2$ |

Table 6. Comparison of solutions

|  | TS | B&D |
| --- | --- | --- |
| Mass | 7062.14 lb | 7064.16 lb |
|  | ~3177.96 kg | ~3178.87 kg |
| Frequency | 28.427 Hz | 28.515 Hz |
| Displacement | 4.38 in | 4.41 in |
|  | ~0.111 m | ~0.112 m |
| Compound objective | 0.4873 | 0.4912 |

Table 7. Comparison of objectives

## CONCLUSIONS

The results presented in this paper show that a variable step size TS can be applied to problems in structural optimisation. The TS method is at least as efficient than many other optimisation methods. The implementation used in this paper is less contrived than other TS implementations but is locating high quality solutions when searching for a single objective solution.

When the TS is implemented using a compounded objective function to search for a multiobjective solution, the search is locating a high quality solution. Direct comparison to the solution found by Bennage and Dhingra (1993) is difficult due to slight numerical differences in the model that cause that solution to violate constraints. These numerical differences may also effect the single objective solutions used by the game theory objective function to calculate the relative performance of the solution away from the best and worst solutions. However, by comparing the actual objective function values it can be seen that the solution found by the TS is not dominated by the solution found by Bennage and Dhingra (1993) and has more desirable values for two of the objective functions.

The use of TS has many merits including ease of implementation and the ability to cope easily with discrete parameters. Future work will continue to refine and enhance the current TS implementation by the investigation of concepts such as frequency, as opposed to recency, based memory and strategic intensification and diversification strategies with the aim of improving the effectiveness and efficiency of the method. To improve performance on problems with multiple objectives a true multiobjective implementation will be developed.